\documentclass[review]{elsarticle}

\usepackage{chemformula} 
\usepackage[T1]{fontenc} 
\usepackage{verbatim}
\usepackage{graphicx}
\usepackage{amssymb}
\usepackage{amsmath}
\usepackage{xspace}
\usepackage{amstext}
 \usepackage{natbib}
 \usepackage{mciteplus}
 \usepackage{lineno}
\usepackage[colorlinks,citecolor=blue,linkcolor=red]{hyperref}
\usepackage{hhline}
\usepackage{multirow}

 \modulolinenumbers[5]

\newcommand{\red}{\textcolor{red}}

\journal{Mater. Today Phys.}
\begin{document}

\begin{frontmatter}

\title {Residual surface charge mediated near-field radiative energy transfer: A topological insulator analog}

\author[mymainaddress,mysecondaddress]{Minggang Luo}
\author[mythirdaddress]{Jiaqi Zhu}
\author[myforthaddress]{S.-A. Biehs\corref{mycorrespondingauthor}}
\cortext[mycorrespondingauthor]{Corresponding author}
\ead{s.age.biehs@uni-oldenburg.de}

\author[mymainaddress,myfifthaddress]{Junming Zhao\corref{mycorrespondingauthor}}
\ead{jmzhao@hit.edu.cn}

\author[mysixthaddress]{Linhua Liu}

\address[mymainaddress]{School of Energy Science and Engineering, Harbin Institute of Technology, 92 West Street, Harbin 150001, China}
\address[mysecondaddress]{Laboratoire Charles Coulomb (L2C) UMR 5221 CNRS-Universit\'e de Montpellier, F- 34095 Montpellier, France}
\address[mythirdaddress]{Center for  Composite Materials and Structures, Harbin Institute of Technology, Harbin 150080, China}
\address[myforthaddress]{Institut f{\"u}r Physik,Carl von Ossietzky Universit{\"a}t, D-26111 Oldenburg, Germany}
\address[myfifthaddress]{Key Laboratory of Aerospace Thermophysics, Ministry of Industry and Information Technology, Harbin 150001, China}
\address[mysixthaddress]{School of Energy and Power Engineering, Shandong University, Qingdao 266237, China}

\begin{abstract}
We study the modifications of near-field radiative energy transfer (NFRET) caused by residual surface charges, which are common in micro- and nano-systems like NEMS/MEMS. The host object with the residual surface charges and the inherent bulk state can be treated as an analog of the real three-dimensional topological insulator, which is inherent of also both surface states and bulk states and is promising to modulate NFRET. Through constructing such a topological insulator analog, we aim to modulate NFRET concerning only common trivial materials. Besides the well-known resonant modes (surface polariton and localized surface polariton) supported by the bulk state, the residual surface charges give rise to an additional temperature-dependent mode providing a new heat flux channel. For low temperatures we find a giant surface-charge-induced enhancement of the NFRET due to a good match between the surface-charge-induced resonance and the Planck window. However, for relative high temperatures where the Fr\"{o}hlich resonance dominates the heat transfer rather the surface-charge-induced resonance, the residual charges result in a weakening of the NFRET. This work paves way for understanding and modulating the near-field radiative energy transfer for micro- and nano-systems.
\end{abstract}


\end{frontmatter}


\section{Introduction}
\begin{figure} [htbp]
\centerline {\includegraphics[width=0.7\textwidth]{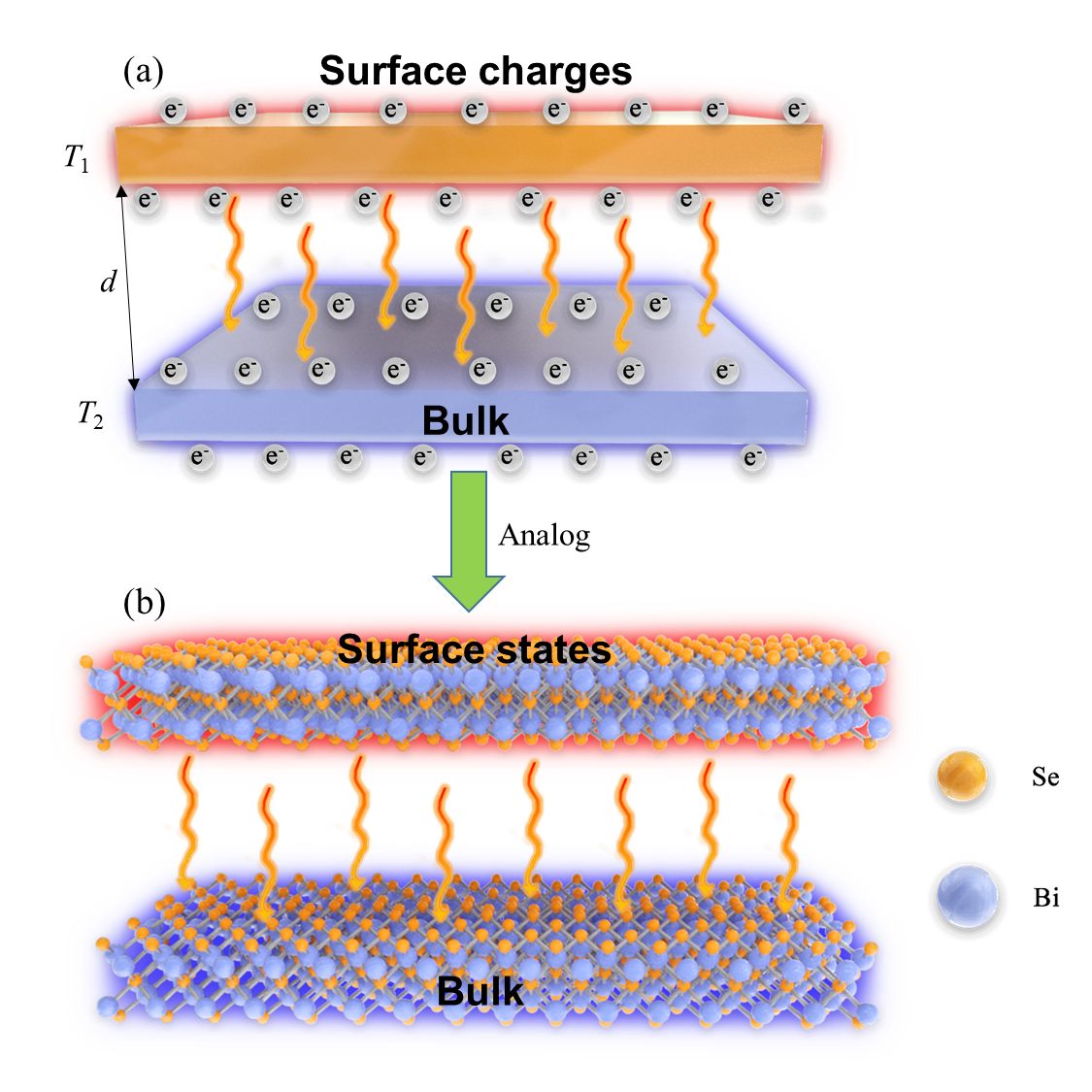}}
\vspace{-1em}
\caption{Sketch of the considered NFRET geometries: (a) two charged objects, where the residual charges are deposited on the surfaces, and (b) the real three-dimensional topological insulator (3D TI), e.g., Bi$_2$Se$_3$. The charged object is an analog to the topological insulator.}
\vspace{-1em}
\label{structure}
\end{figure}

The energy transfer between two objects with a separation distance comparable to or less than the thermal wavelength $\lambda_T=\hbar c/k_BT$ via thermal photons can exceed by several orders of magnitude the blackbody limit, which has been investigated widely from both theoretical side and experimental side for many different geometrical configurations (e.g., planar structures \cite{Rytov1989,Polder1971,Loomis1994,Carminati1999,Volokitin2001,Ottens2011,Watjen2016,Ghashami2018,DeSutter2019} and nanoparticles-involved structures \cite{Chapuis2008,Narayanaswamy2008,Manjavacas2012,Chapuis2008plate,Messina2018,DongPrb2018,Shen2009,Rousseau2009,Song2015}) as reviewed in Ref.~[\citenum{SABEtAl2021}], for instance. In the aforementioned works, only electrically neutral objects are involved. However, electrically neutral objects can easily be charged by exchanging electrons or ions with their environments~\cite{Bohren1977,Spitzer1998,Vladimirov2005,Kocifaj2020}. As shown in Fig.~\ref{structure} (a), the residual charges are localized on the surface of the object  \cite{Klacka2007,Klacka2015}. Hence, the charged object of both the residual surface charges and the inherent bulk states is much like the topological insulator (e.g., Bi$_2$Se$_3$), which inherently prosesses a topological surface state and bulk state, as shown in Fig.~\ref{structure} (b). Recently, modulations of the near-field radiative heat transfer for the real three-dimensional \cite{Wu2022MTP} and two-dimensional \cite{Liu20211iSCI} topological insulators Bi$_2$Se$_3$ were investigated, respectively. For the three-dimensional topological insulator (3D TI) Bi$_2$Se$_3$, the coupling between the Dirac plasmons supported by the non-trivial surface states and the multiple phonon polariton modes supported by the bulk states is found to be able to strongly modulate and enhance the near-field radiative energy transfer \cite{Wu2022MTP}. In addition to the bulk and non-trivial surface states, the edge states are demonstrated to be another tool for modulation of NFRET \cite{Tang2019PRApplied}. For sure, the real 3D TI Bi$_2$Se$_3$ is so special in its optical properties that it has such promising heat transfer modulation properties. For a more general purpose, we are expecting to construct a topological insulator analog composed of the common materials easily found in practice with the help of the surface localized residual charges, to modulate heat transfer between common materials of only intrinsic trivial states. 

The presence of residual charges significantly affects the optical properties of nanostructures, e.g.\ the absorption and scattering cross sections of nanoparticles~\cite{Kocifaj2020,Klacka2007,Klacka2015,Zhang2020particle}, the reflection and transmission coefficients of planar structures~\cite{Zhang2020slab} in the infrared region. Since the optical properties of nanostructures in the infrared determine their ability to exchange heat with their environment by thermal radiation, it can be expected that thermal radiation of charged objects in general and near-field radiative energy transfer (NFRET) in particular will be affected by the presence of residual charges.  In the extreme near field, the effects of the electrostatics on the heat transfer have already been studied  \cite{Volokitin2019JETP,Volokitin2021prb,Takuro2021prb,Guo2022prb}. However, due to the lack of studies on this topic in the near field rather than extreme near field where the coulomb interaction becomes relatively weak, it is largely unknown how the presence of residual charges will affect NFRET in different geometries.

We address the aforementioned missing points in this paper, investigating the NFRET transfer between charged objects, e.g.\ charged nanoparticles and charged planar slabs within the framework of fluctuational electrodynamics~\cite{Rytov1989,Polder1971,Ben2011}, by introducing the residual charge modification. This work is organized as follows. In Sec.~\ref{theory}, the residual charge conductivity model and its modification to the optical properties used in the fluctuational electrodynamics for the charged objects are presented briefly. In Sec.~\ref{Results}, we will analyze that the residual charges effect and its temperature dependence on the near-field thermal radiation for the considered two typical structures (particle and planar structures),  as well as the behind mechanism.

\section{Theory}
\label{theory}
The real topological insulators have both the insulating bulk states and the topologically protected conductive surface states, of which the optical properties are characterized by the two-dimensional conductivity \cite{Wu2022MTP}. Similarly, the residual charges on the topological insulator analog (i.e., charged object) are usually localized at its surface\cite{Bohren1977,Kocifaj2020,Zhang2020slab}, which also result in a surface conductivity defined as follows \cite{Klacka2007}.
\begin{equation}
  \sigma_{s}=\frac{\eta e/m_e}{k_{\rm{B}} T/\hbar-i\omega},
  \label{Sigma_s}
\end{equation}
where $\eta$ is the surface charge density, $e$ is the electron charge, $m_e$ is the electron mass, $k_{\rm{B}}$ is the Boltzmann constant, and $\hbar$ is the reduced Planck constant. For the charged slabs, the surface charge density is $\eta = e N$ where $N$ [m$^{-2}$] is the surface charge number density~\cite{Zhang2020slab} while for charged nanoparticles of the radius $a$ we have $\eta = \epsilon_0 \phi / a$ where $\phi$ [V] is the surface potential~\cite{Klacka2007,Zhang2020particle} and $\epsilon_0$ is the vacuum permittivity. From Eq.~(\ref{Sigma_s}) it can be seen that the surface conductivity induced by the residual charges is temperature-dependent. In particular, the surface conductivity decreases when increasing the temperature, because the charge motion and collisions become more significant in that case.

When considering the NFRET between the charged objects, like the method that has been used to take into account the topologically protected surface states by introducing a surface conductivity \cite{Wu2022MTP}, we also introduce a similar surface conductivity but of the residual charges origin in the fluctuational electrodynamics. In the following, we will show the surface-charge-induced modifications to the classical fluctuational electrodynamics for neutral objects for different configurations, i.e., small particles and planar slabs, respectively.

Let's start by considering NFRET between two identical charged nanoparticles. The expression for the exchanged power $P$ due to NFRET between a particle $1$ and particle $2$ within the framework of fluctuational electrodynamics for small particles within the dipole approximation can be written in a Landauer form as~\cite{Ben2011,DongPrb2017,Luo2019}
\begin{equation}
  P = 3 \int_{0}^{+\infty} \frac{\mathrm{d}\omega}{2\pi}\left[\Theta(\omega,T_1)-\Theta(\omega,T_2)\right]\mathcal{T}_{12}(\omega),
  \label{ExchangedPower}
\end{equation}
where $\omega$ is the angular frequency, $\Theta(\omega,T) = \hbar \omega / (\exp(\hbar\omega/k_{\rm B} T) - 1)$ is the mean energy of the harmonic oscillator at temperature $T$ and $\mathcal{T}_{12}(\omega)$ is the transmission coefficient between the two particles given as follows~\cite{Ben2011,DongPrb2017,Luo2019}.
\begin{equation}
\begin{aligned}
\mathcal{T}_{12}(\omega)=&\frac{4}{3}k^4\left[\textrm{Im}(\chi_E^1)\textrm{Im}(\chi_E^2)\textrm{Tr}(G_{12}^{EE}G_{12}^{EE\dagger})\right.\\&+\textrm{Im}(\chi_E^1)\textrm{Im}(\chi_H^2)\textrm{Tr}(G_{12}^{EM}G_{12}^{EM\dagger})\\&+\textrm{Im}(\chi_H^1)\textrm{Im}(\chi_E^2)\textrm{Tr}(G_{12}^{ME}G_{12}^{ME\dagger})\\&+\left.\textrm{Im}(\chi_H^1)\textrm{Im}(\chi_H^2)\textrm{Tr}(G_{12}^{MM}G_{12}^{MM\dagger})\right],
\label{transmission}
 \end{aligned}
\end{equation}
where the parameters $\chi_E^{}=\alpha_E^{}-\frac{ik^3}{6\pi}
\left|\alpha_E^{}\right|^2$, $\chi_H^{}=\alpha_H^{}-\frac{ik^3}{6\pi}
\left|\alpha_H^{}\right|^2$, $k$ is the wavevector in vacuum, $G_{ij}^{\nu\tau}$ ($\nu$,$\tau$=$E$~or~$M$) is the Green\textquotesingle s function for the many-particle system. The transmission coefficient $\mathcal{T}_{12}(\omega)$ depends on the optical properties of the two particles, i.e.\ their electric and magnetic dipolar polarizabilities ${{\alpha }_{E}}=\frac{i6\pi }{{{k}^{3}_0}}{{a}_{1}}$ and ${{\alpha }_{H}}=\frac{i6\pi }{{{k}^{3}_0}}{{b}_{1}}$, where $k_0 = \omega/c$ with the vacuum light velocity $c$, $a_1$ and $b_1$ are the dipolar scattering coefficients. Due to the presence of residual charges these scattering coefficients derived from Mie theory are modified such that~\cite{Klacka2007,Zhang2020particle}
\begin{align}
	a_1 &=\frac{\sqrt{\varepsilon} \Psi_1(y)\Psi_1'(x)- \Psi_1(x)\Psi_1'(y)+ \frac{i\mu_0\omega\red{\sigma_s}}{k_0} \Psi_1'(x) \Psi_1' (y)}{\sqrt{\varepsilon} \Psi_1 (y) \xi_1'(x)- \xi_1 (x) \Psi_1'(y) + \frac{i\mu_0\omega\red{\sigma_s}}{k_0} \xi_1'(x) \Psi_1'(y)},
\label{a1} \\
	b_1 &=\frac{\Psi_1(y)\Psi_1'(x)- \sqrt{\varepsilon} \Psi_1 (x) \Psi_1'(y) +\frac{i\mu_0\omega\red{\sigma_s}}{k_0} \Psi_1(x) \Psi_1 (y)}{\Psi_1(y) \xi_1'(x) -  \sqrt{\varepsilon} \xi_1 (x)\Psi'(y) + \frac{i\mu_0\omega\red{\sigma_s}}{k_0} \xi_1(x) \Psi_1(y)},
\label{b1}
\end{align}
where $x=k_0 a$, $y=\sqrt{\varepsilon }k_0a$, $\varepsilon$ is the relative dielectric permittivity, $\mu_0$ is the vacuum permeability, $\Psi_1(x) = x j_1(x)$, $\xi_1(x) =  x h_1^{(1)}(x)$ are the Riccatti Bessel functions and ${{j}_{\text{1}}}(x)$, $h_{\text{1}}^{(1)}(x)$ are the spherical Bessel and Hankel functions. For the neutral particles ($\sigma_s=0$), the coefficients $a_1$ and $b_1$ defined by the Eqs.~(\ref{a1}) and (\ref{b1}) reduce to the well-known Mie scattering coefficients~\cite{Bohren1983}. 

Similarly, for two identical charged planar slabs with the surface conductivity $\sigma_s$ fluctuational electrodynamics allows to determine the flux $\Phi$ [W/m$^2$] between those slabs yielding again a Landauer form~\cite{Rytov1989,Polder1971,SAB2010} 
\begin{equation}
	{\Phi}=\int_{0}^{\infty} \!\! \frac{\text{d}\omega}{2 \pi} \, \left[ \Theta (\omega ,{{T}_{1}})-\Theta (\omega ,{{T}_{2}}) \right] \int_{0}^{\infty } \!\! \frac{\text{d}\beta}{2 \pi} \, \beta \sum_{i = {\rm s}, {\rm p}}\mathcal{T}_i(\omega ,\beta ),
\label{P_slab}
\end{equation}
where $\beta$ is the magnitude of the in-plane wave vector and the $\mathcal{T}_i(\omega ,\beta )$ ($i = {\rm s}, {\rm p}$) is the energy transmission coefficient for the s- and p-polarized waves, which again strongly depends on the optical properties of the two slabs via their reflection coefficients $r_{i}$ ($i = {\rm s}, {\rm p}$). Hence, the impact of residual surface charges modifies the heat flux by modifying the reflection as follows \cite{Zhang2020slab,Shi2017Acs,Zhao2017hBN} 
\begin{align}
  r_{\rm s}&=\frac{k_z^{(0)}-k_z-\red{\sigma_s}\mu_0\omega}{k_z^{(0)}+k_z+\red{\sigma_s}\mu_0\omega}, \label{r_s} \\
  r_{\rm p}&=\frac{k_z^{(0)}\varepsilon_{}-k_z\varepsilon_{}^{(0)}+\red{\sigma_s} k_z^{(0)}k_z\varepsilon_0^{-1}\omega^{-1}}{k_z^{(0)}\varepsilon_{}+k_z\varepsilon_{}^{(0)}+\red{\sigma_s} k_z^{(0)}k_z\varepsilon_0^{-1}\omega^{-1}},
  \label{r_p}
\end{align}
where $k_z^{(0)}=\sqrt{\varepsilon_{}^{(0)}k_0^{2}-\beta^{2}}$ and  $k_z=\sqrt{\varepsilon k_0^{2}-\beta^{2}}$. For the neutral slabs ($\sigma_0=0$), the above reflection coefficients Eqs.~(\ref{r_s}) and (\ref{r_p}) reduce to the well-known Fresnel reflection coefficients~\cite{Volokitin2001,Shi2017Acs}. The energy transmission coefficient $\mathcal{T}_i(\omega ,\beta )$ ($i = {\rm s}, {\rm p}$) is defined as follows \cite{Wu2022MTP,Shi2017Acs}.
\begin{equation}
\mathcal{T}_i(\omega ,\beta )=\left\{
\begin{array}{rcl}
\frac{(1-|r_{i}^{1}|^2)(1-|r_{i}^{2}|^2)}{|1-r_{i}^{1}r_{i}^{2}e^{2ik_z^{(0)}d}|^2},  & {\beta <k_0}\\
\frac{4{\rm Im}(r_{i}^{1}){\rm Im}(r_{i}^{2})e^{2ik_z^{(0)}d}}{|1-r_{i}^{1}r_{i}^{2}e^{2ik_z^{(0)}d}|^2},  & {\beta >k_0}
\end{array}
\right.
\label{Transmission}
\end{equation}
where $r_{i}^{1}$ and $r_{i}^{2}$ are $i$-polarized reflection coefficient for slab 1 and 2, respectively.

Here we assume that the conductance inside the materials is much larger than the radiative conductance between the objects so that temperature gradients inside the materials are small can be neglected. This is indeed well fulfilled for the materials considered in our manuscript. A treatment of the impact of temperature gradients can be found in Ref.~\cite{Reina2020prl}.

\section{Results and discussion}
\label{Results}
In order to study the impact of residual charges on the exchanged heat between two topological insulator analogs (particle-type and slab-type, respectively), we choose for the numerical calculation the typical dielectric material SiC for both nanoparticles and slabs. The dielectric functions of the bulk SiC state is described by the Lorentz-Drude model  $\varepsilon(\omega) =\varepsilon_{\infty}^{}(\omega^2-\omega_l^2+i\gamma\omega)/(\omega^2-\omega_t^2+i\gamma\omega)$ with $\varepsilon_{\infty}^{}$ = 6.7, $\omega_l^{}$ = 1.827 $\times$ 10$^{14}$ rad$\cdot$s$^{-1}$, $\omega_t^{}$ = 1.495 $\times$ 10$^{14}$ rad$\cdot$s$^{-1}$, and $\gamma$ = 0.9 $\times$ 10$^{12}$ rad$\cdot$s$^{-1}$ \cite{Palik}. The particle radius will be set to $a=50$ nm. The separation edge to edge between two nanoparticles or slabs is labeled by $d$. The minimum $d_{\rm{min}}$ between the nanoparticles is 2$a$ guaranteeing the validity of the dipole approximation~\cite{SABEtAl2021,DongPrb2017}. Below, we will separately investigate and discuss the characteristics of the NFRET for the particle-type and slab-type topological insulators.

\subsection{NFRET between two particles with residual charges}

\begin{figure} [htbp]
\centerline {\includegraphics[width=1.\textwidth]{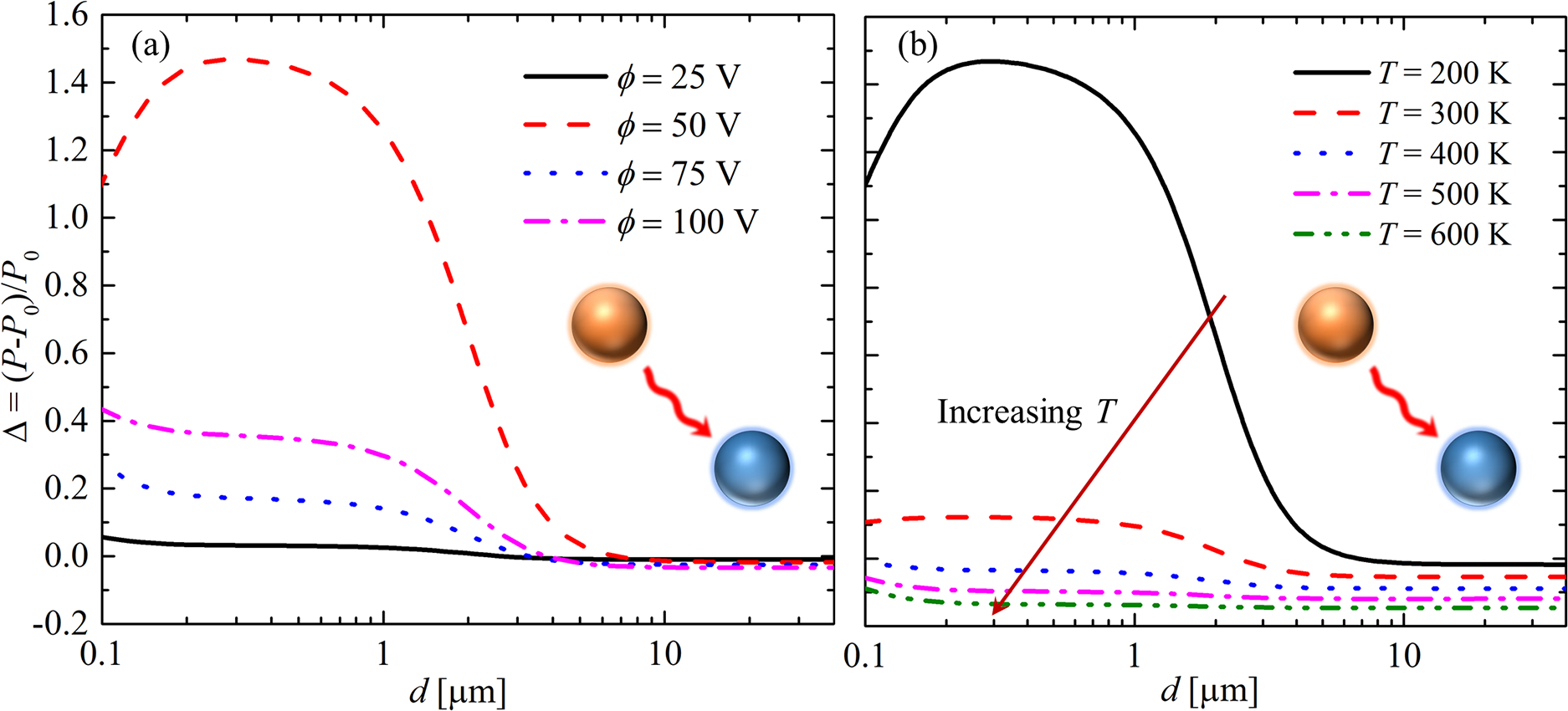}}
        \caption{Dependence of the $(P-P_0)/P_0$ on the separation $d$ between two nanoparticles. $P$ is radiative heat flux with residual charges, $P_0$ is the radiative heat flux with no charge. (a) The two particles are fixed at 200 K and 0 K. $\phi=$25 V, 50 V, 75 V and 100 V, respectively. The considered surface potentials are comparable to these used in the Refs.~[\citenum{Klacka2007,Zhang2020particle}]. (b) $\phi=50$ V. One particle is fixed at 0 K. The temperature of the other particle takes the following values, $T=$200 K, 300 K, 400 K, 500 K and 600 K.}
        \label{charge_effect_NP}
\end{figure}

To study the impact of residual charges we first define the relative deviation $\Delta = (P-P_0)/P_0$ of the exchanged power $P$ with residual charges with respect to the exchanged power $P_0$ without residual charges. The dependence of $\Delta$ on the separation $d$ between the two charged nanoparticles is shown in Figs.~\ref{charge_effect_NP} (a) and (b). In Figs.~\ref{charge_effect_NP} (a),  the two particles are fixed at 200 K (emitter) and 0 K (receiver) and the voltage (or amount of residual charges) is varied. Following the convention used in Refs.~\cite{Ben2011,DongPrb2017}, the receiver is fixed at 0 K to study the emission purely given by the emitter. The surface potentials used in this work are comparable to that of the Refs.~[\citenum{Klacka2007,Zhang2020particle}]. It can be seen that in the near-field regime with separations $d < 4~\mu$m there is a clear enhancement of the exchanged power up to about $2.5 P_0$ when increasing the amount of residual charges. That there is only a strong residual charge effect for such small distances is due to fact that the residual charges are strongly localized at the particle surfaces having a strongly decaying electric field. In Figs.~\ref{charge_effect_NP} (b), again one particle (receiver) is fixed at 0 K and temperature of the other particle (emitter) is varied from 200 K to 600 K choosing the surface potential $\phi = 50\,{\rm V}$. It can be observed that for low temperatures the enhancement due to the residual charges is strong. When increasing the temperature $\Delta$ gets smaller. For temperatures larger than 200K there is also a sign change at a given distance. When increasing the temperature this distance moves to smaller values such that at 400K we find negative $\Delta$ for all shown distance. Hence, for relatively large temperature the exchanged power is diminished by the presence of the residual charges.

\begin{figure} [htbp]
  \centerline{\includegraphics[width=1.\textwidth]{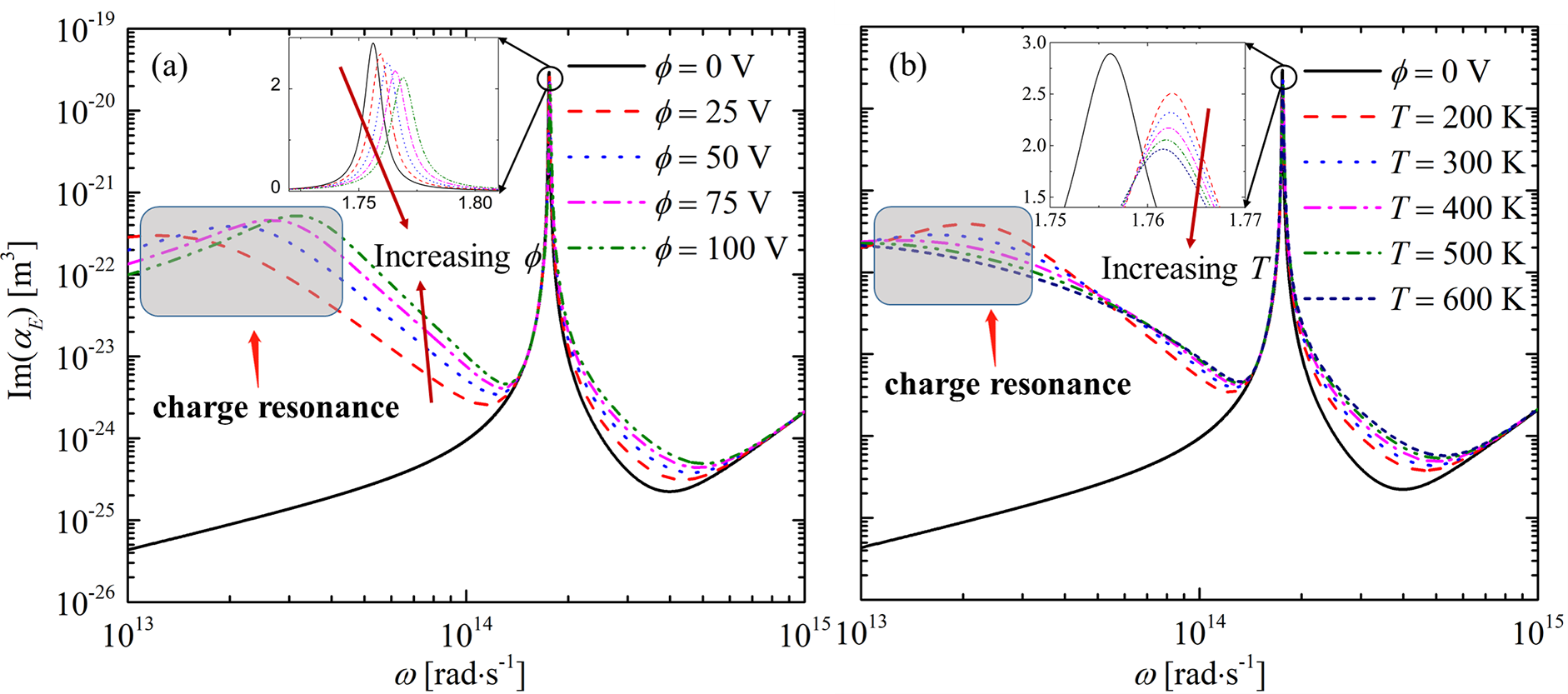}}
  \caption{Charged SiC nanoparticle ``absorptivity'' ${\rm Im}(\alpha_{\rm E})$: (a) for different potential (i.e., 25 V, 50 V, 75 V and 100 V) at $T=200$ K and (b) at different temperatures (i.e., $T=200$ K, 300 K, 400 K, 500 K and 600 K) and fixed potential $\phi=50$ V. The polarizability of the neutral SiC nanoparticle ($\phi=0$) is shown as reference. The region around the Fr{\"ohlich} resonance peak  at $\omega_{\rm r} = 1.756\times10^{14}\,{\rm rad/s}$ is enlarged. }
   \label{polarizability}
\end{figure}

Since the NFRET between the nanoparticles strongly depends on their absorptivities which are proportional to ${\rm Im}(\alpha_{\rm E})$~\cite{Bohren1983}, we can get some insight in the underlying mechanism behind the observed surface-charge-induced effects by focusing on ${\rm Im}(\alpha_{\rm E})$ of the SiC nanoparticles as shown in Fig.~\ref{polarizability}. Besides the well-known Fr{\"ohlich} resonance at $\omega_{\rm r} = 1.756\times10^{14}\,{\rm rad/s}$ from the condition ${\rm Re}[\varepsilon(\omega_{\rm r})] = - 2$ supported by the bulk state, a new surface-charge-induced resonance can be observed for $\omega < \omega_{\rm r}$. This resonance is consistent with the reported surface-charge-induced resonance around wavelength of 1 $\mu$m of SiO$_2$ and some other particles~\cite{Kocifaj2015resonance}. Due to the fact, that this resonance has such a small frequency, it is clear that it has a particularly large impact at low temperatures. As shown in the Fig.~\ref{polarizability}(a), by increasing the potential, the Fröhlich peak shows a blueshift and decreases slightly. However, the surface-charge-induced peak shows a blueshift and increases slightly when increasing the potential. The Fröhlich resonance and surface-charge-induced resonance work together and result in a non-monotonic behavior when increasing the potential, as shown Fig.~\ref{charge_effect_NP}(a). To understand now, why for large temperatures $\Delta$ can change its sign as observed in Fig.~\ref{charge_effect_NP}(b), we show in Fig.~\ref{polarizability} (b) a close-up of the Fr\"{o}hlich resonance peak. Obviously, the resonance peak is blue shifted due to the presence of the residual charges and by increasing the temperature it is becoming weaker. Hence, for larger temperatures where the Fr\"{o}hlich resonance dominates the heat transfer rather than the surface-charge-induced resonance, we find a smaller NFRET $P$ when the particles are charged compared to the neutral case $P_0$. By increasing the temperature $P$ becomes smaller and smaller compared to $P_0$ as observed in Fig.~\ref{charge_effect_NP}(b).

\subsection{NFRET between two slabs with residual charges}
From our analysis of the NFRET between particle-type topological insulator analogs, we can expect that there is also an enhancement of NFRET for charged slabs (slab-type topological insulator analogs) which might be particularly significant at low temperatures. Therefore, we will now focus on the NFRET between two charged slabs at 10K so that the Planck window matches well with the spectral position of the surface-charge-induced resonance~\cite{Kocifaj2015resonance}. We note, that the NFRET between two neutral slabs has already been experimentally studied at the such low temperatures, recently~\cite{Kralik2012prl,Kralik2021}. Instead of studying the heat flux as epressed in Eq.~(\ref{P_slab}) we focus on the heat transfer coefficient (HTC, $h$ [W/(m$^2 \cdot$K)]) defined by 
\begin{equation}
  h=\lim\limits_{T_1 \rightarrow T_2} \frac{\Phi}{T_1-T_2}=\int_0^{\infty}h_\omega {\rm d}\omega, 
\end{equation}
where $h_\omega$ [W/(m$^2 \cdot$K$\cdot$rad$\cdot$s$^{-1}$)] is the spectral HTC. 

\begin{figure} [htbp]
\centerline {\includegraphics[width=.9\textwidth]{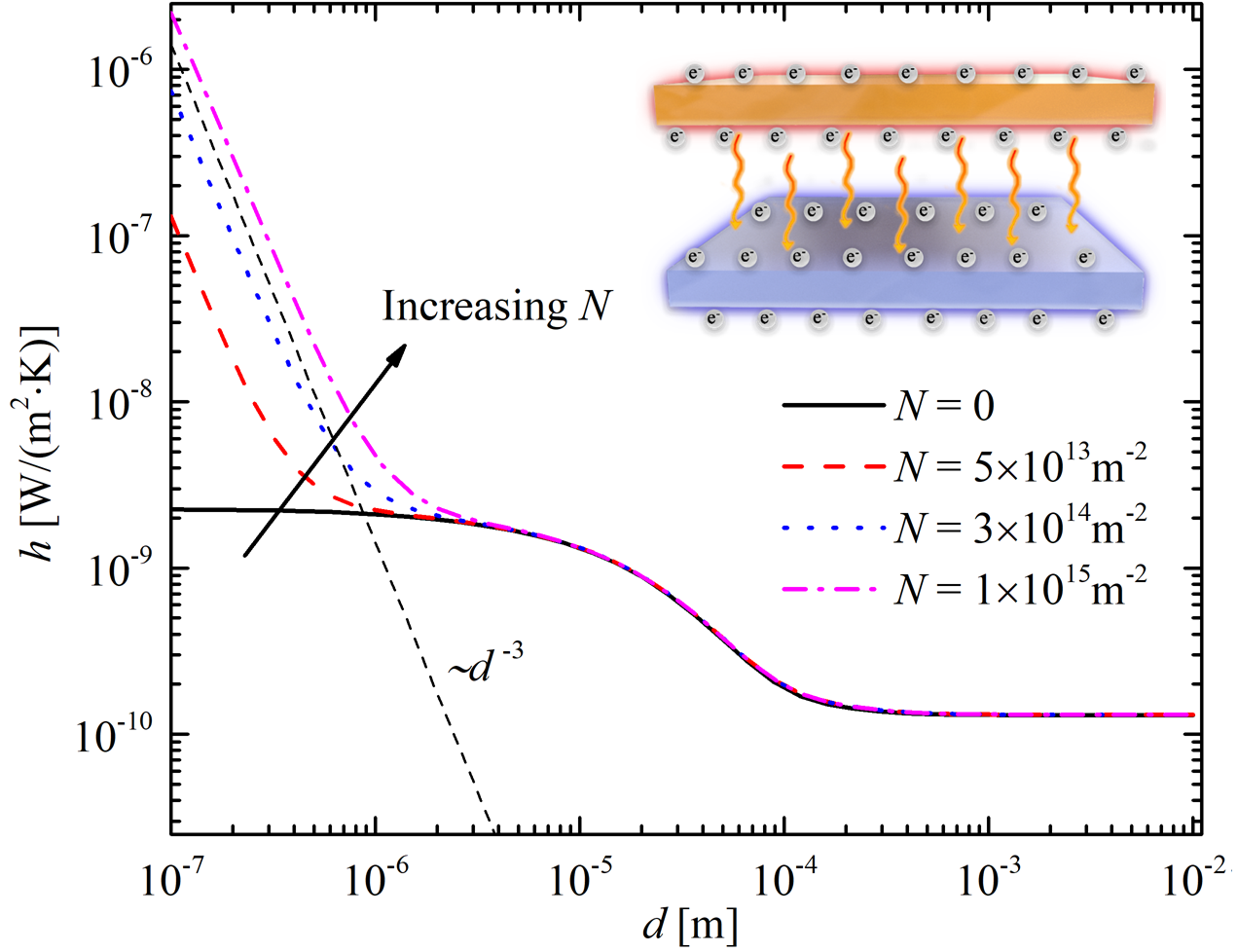}}
        \caption{Distance dependence of the HTC $h$ between two charged SiC slabs at 10 K considering the number densities $N=0,~5\times10^{13}$m$^{-2},~3\times10^{14}$m$^{-2}$ and $1\times10^{15}$m$^{-2}$. The line of $1/d^{3}$ is added for reference.}
        \label{charge_slab_Q}
\end{figure}

In Fig.~\ref{charge_slab_Q}, we show the full HTC as function of separation distance $d$ for different surface number densities $N$. It can be seen that due to the presence of surface charges there is a strong near-field enhancement of about two to three orders of magnitude for $d = 100\,{\rm nm}$ and the chosen values of $N$. This observation is radically different from the negligible electrostatic effect for the metal planar structures at room temperature reported recently in Ref.~[\citenum{Guo2022prb}]. We find numerically that the heat flux induced by the presence of the surface charges is approximately proportional to 1/$d^3$ in the shown distance regime, so that at distances below 100nm the enhancement would be even much larger. However, it has to be kept in mind that the charges also induce a force on the plates and in addition if the gap between the slabs is not filled with vacuum but a gas, for instance, there can be a discharge when the voltage reaches the breakdown voltage of the gas. For the largest values of $N = 10^{15}\,{\rm m}^{-2}$ and $3\times10^{14}\,{\rm m}^{-2}$ we find a presure of $F/A = 1.4\,{\rm kPa}$ and $130\,{\rm Pa}$  which is about 100 or 10 times the Casimir force at 100nm. For the voltage we find $1.8\,{\rm V}$ or $0.5\,{\rm V}$ at $100\,{\rm nm}$ so that the value corresponds to 6 or 1.7 times the breakdown voltage of air. Hence, values of $N > 5\times10^{13}\,{\rm m}^{-2}$ can only be achieved under vacuum conditions as typically used in near-field experiments~\cite{Kralik2012prl,Kralik2021}.

\begin{figure} [htbp]
\centerline {\includegraphics[width=0.8\textwidth]{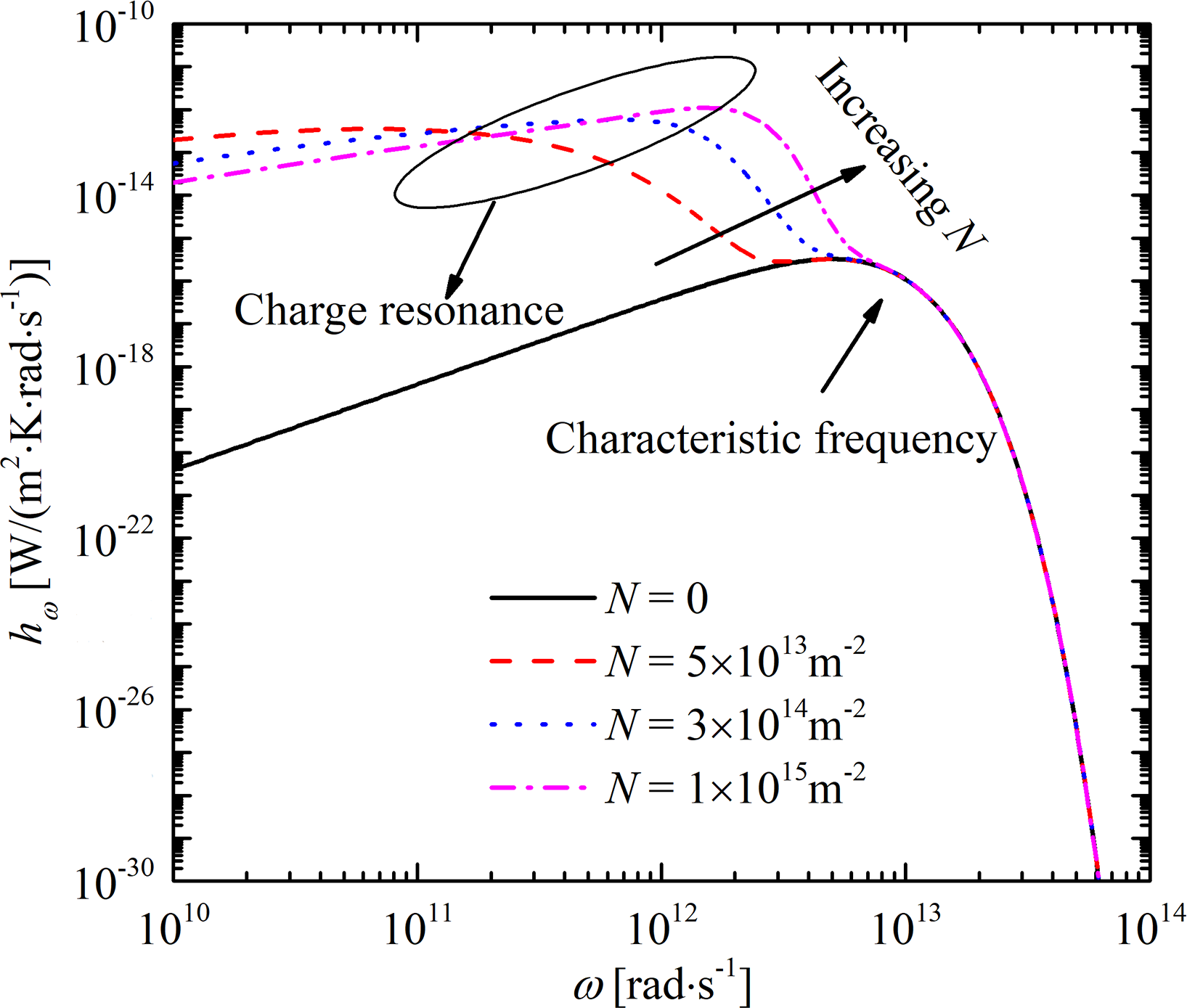}}
	\caption{Spectrum of the HTC $h_\omega$ at $T=10$~K for $N=0,~5\times10^{13}$m$^{-2},~3\times10^{14}$m$^{-2}$ and $1\times10^{15}$m$^{-2}$ chosing the separation distance $d = 100\,{\rm nm}$.}
        \label{spectral_slab_Q}
\end{figure}

To understand the mechanism behind the enhancement of HTC, we show the corresponding spectral HTCs in Fig.~\ref{spectral_slab_Q}. Two kind of peaks can be observed: (1) a broad peak due to the Planck spectrum (basically shonw by the curve for $N = 0$) which is independent of the charge density; (2) a low frequency peak corresponding to the surface-charge-induced resonance. Obviously, when increasing the charge number density this peak enhances the HTC by shifting towards the Planck window. This can be explained by increased charge collisions resulting in a blueshift of the surface-charge-induced resonance. Similarly to the case of two nanoparticles, when the separation $d$ increases the surface-charge-induced enhancement effect decreases dramatically, which can again be attributed to the fact that the resonant mode due to residual charges is highly localized on the surfaces of the slabs. Note, that the surface phonon polariton resonance of SiC is at $\omega = 1.787\times10^{14}\,{\rm rad/s}$ and therefore cannot be thermally excited at 10~K. This also explains, why there the HTC does not show any $1/d^2$ dependence associated with the surface phonon polaritons in Fig.~\ref{charge_slab_Q}.

\begin{figure} [htbp]
    \centerline {\includegraphics[width=1.\textwidth]{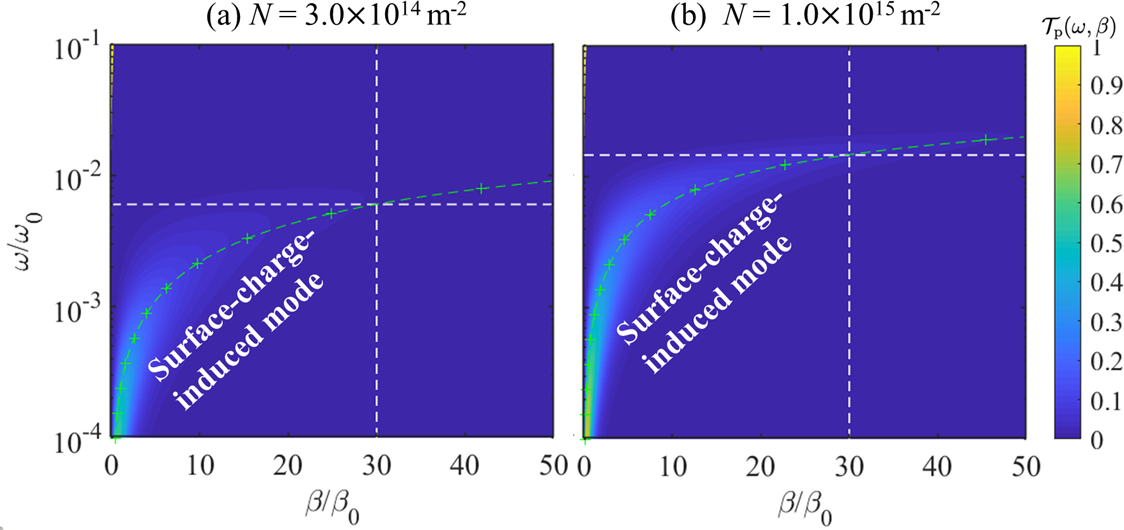}}
	\caption{The transmission coefficient $\mathcal{T}_{\rm p}(\omega,\beta)$ for $p-$polarized waves for two SiC slabs at separation $d=100$ nm with a charge number density $N$ given by (a) 3.0$\times 10^{14}$m$^{-2}$ and (b) 1.0$\times 10^{15}$m$^{-2}$. The frequencies and wave vectors are normalized to $\omega_0=$1.0$\times 10^{14}$rad$\cdot$s$^{-1}$ and $\beta_0^{}=\omega_0/c$.}
        \label{transmission_coefficients}
\end{figure}

To further understand the mechanism behind the enhancement of HTC, we show the transmission coefficient $\mathcal{T}_{\rm p}(\omega,\beta)$ of the p-polarized waves for $d =  100\,{\rm nm}$ in Fig.~\ref{transmission_coefficients} for $N=$ 3.0$\times 10^{14}$m$^{-2}$ and 1.0$\times 10^{15}$m$^{-2}$. The enhanced transmission due to the surface-charge-induced mode can be nicely seen , whereas the surface phonon polariton supported by the bulk state is outside the plotted frequency range. When increasing the charge number density [from Fig.~\ref{transmission_coefficients} (a) to Fig.~\ref{transmission_coefficients} (b)], the surface-charge-induced mode blueshifts and approaches the Planck window at 10K. Hence, more and more energy can be transferred between the two slabs when increasing the charge density, which accounts for the increasing enhancement of HTC observed in Fig.~\ref{charge_slab_Q}. In Fig.~\ref{transmission_coefficients}, the dispersion relation of the surface-charge-induced resonance is shown as dashed-symbol line. It is determined by the poles of the reflection coefficient $r_{\rm p}$ in Eq.~(\ref{r_p}). For the strong evanescent limit $\beta \gg k_0$ it is given by 
\begin{equation}
  \beta=\frac{(\varepsilon+1)\varepsilon_0 \omega m_e(i k_B T/\hbar+\omega)}{Ne^2}.
  \label{dispersion_charge_mode}
\end{equation}

\begin{figure} [htbp]
    \centerline {\includegraphics[width=.95\textwidth]{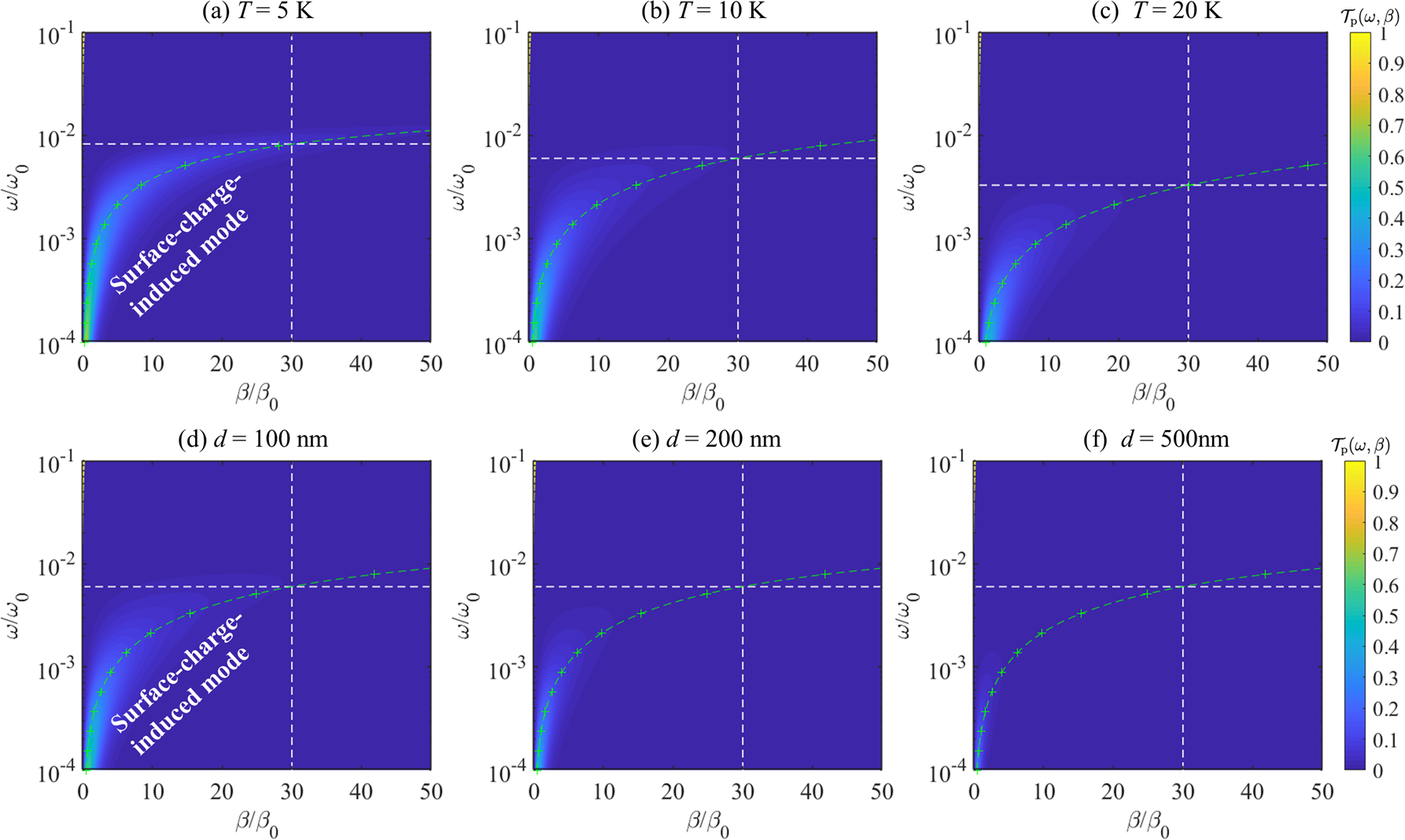}}
    \setlength{\abovecaptionskip}{-0.05cm}
	\caption{The transmission coefficient $\mathcal{T}_{\rm p}(\omega,\beta)$ for $p-$polarized waves for two SiC slabs at separation $d=100$ nm with a charge number density $N$ = 3.0$\times 10^{14}$m$^{-2}$ at different temperature (a) $T=5$ K, (b) 10 K, and (c) 20 K. The transmission coefficient $\mathcal{T}_{\rm p}(\omega,\beta)$ at 10 K with a charge number density $N$ = 3.0$\times 10^{14}$m$^{-2}$ at different separation (d) $d=100$ nm, (e) $d=200$ nm, and (f) $d=500$ nm.}
        \label{transmission_coefficients_T_d}
\end{figure}

From this relation and the Fig.~\ref{transmission_coefficients} we can easily understand the blueshift of the surface-charge-induced resonance. Because the evanescent waves with $\beta \approx 1/d$ dominate the heat flux between the slabs at a given distance $d$~\cite{SAB2010} we could chose a fixed value of $\beta = 1/d$ in the dispersion relation in Eq.~(\ref{dispersion_charge_mode}). It is then clear that when increasing $N$ the numerator has to increase as well to obtain $1/d$ which can be achieved by increasing $\omega$ (assuming that dispersion can be neglected in that frequency range) explaining the observed blueshift in the spectral HTC in Fig.~\ref{spectral_slab_Q}. Likewise, one can think of a vertical line in Fig.~\ref{transmission_coefficients} at $\beta = 1/d$ which is about 30 $\beta_0$ for $d = 100\,{\rm nm}$. The intersection point of that line with the dispersion relation marks the spectral position of the resonance. Then it is obvious from Fig.~\ref{transmission_coefficients} that the resonances are blue-shifted when increasing $N$. In addition to the charge number density $N$, we show the transmission coefficient for $d$ = 100 nm in Figs.~\ref{transmission_coefficients_T_d}(a), (b) and (c) for the temperature $T$ = 5 K, 10 K and 20 K, respectively. The surface conductivity is dependent on the temperature. Thus, the transmission coefficient is also dependent on the temperature. The transmission coefficient decreases significantly with the increasing temperature $T$. We show the transmission coefficient for the separation $d$ = 100 nm, 200 nm, and 500 nm in Figs.~\ref{transmission_coefficients_T_d}(d), (e) and (f). The transmission coefficient also decreases significantly with the increasing separation $d$. That is the surface charge effect on NFRET is surface-localized. The surface-charge-induced mode becomes less important with increasing the temperature and separation.

So far, we can give a short comment on the comparison of the topological insulator analog (residual charged object) and the real topological insulators involved near-field radiative energy transfer, as summarized in the following Table (\ref{summary}). The real topological insulator is inherent of both surface state and bulk states, which helps for the modulation of NFRET. The Dirac plasmons supported by the surface states and the multiple phonon polariton modes supported by the bulk states not always have strong coupling to help for NFRET. However, the two states even can compete with each other, e.g., (1) decreasing the thickness will enhance the surface state contribution but suppress the bulk state contribution, and (2) increasing the chemical potential will inhibit the phonon polariton modes of the bulk states but enhance the Dirac plasmon of surface states. As a result of the competition between the two states when increasing the thickness, the radiative heat flux between two thin Bi$_2$Se$_3$ films can even be much larger than that between two thick Bi$_2$Se$_3$ films, where the single Dirac plasmon of the surface state leads to a significant heat flux without existence of the multiple phonon polariton modes supported by the bulk states. From the spectrum point of view, as shown in Fig. 8 of Ref.~[\citenum{Wu2022MTP}], the Dirac plasmon of the surface state and the multiple phonon polariton of the bulk states are coupling together in a short angular frequency range. However, as shown in Figs.~\ref{polarizability},~\ref{spectral_slab_Q} and~\ref{transmission_coefficients}, the surface-charge-induced mode is quite far away from the surface phonon polariton. To be more exact, in Figs.~\ref{spectral_slab_Q} and~\ref{transmission_coefficients}, the the surface phonon resonance is even located out of the listed angular frequency. The perfect match between the surface-charge-induced mode and the Planck window (as shown in Fig.~\ref{spectral_slab_Q}) will significantly enhance NFRET by several orders of magnitudes, as shown in Fig.~\ref{charge_slab_Q}.
 However, the mismatch between the surface-charge-induced surface mode and the Planck window (as shown in Figs.~\ref{polarizability}) accounts for a weak enhancement of NFRET, as shown in Fig.~\ref{charge_effect_NP}. Thus, the residual charge on an object can also provide a wide modulation range of NFRET. The topological insulator in practice is not very common in fact. However, one can much more easily construct a topological insulator analog by applying the charges on the common materials (e.g., SiC) to provide an alternative way to modulate the NFRET between the common materials.

\vspace{1em}
\begin{table}[htbp]
  \caption{Comparison of the topological insulator analog and real topological insulators involved near-field radiative energy transfer}
  \begin{tabular}{|l|l|l|}
   \hline
    \multirow{2}*{Items} & Topological insulator analog  & Real topological \\
      & (Residual charged objects) & insulators (Ref.~[\citenum{Wu2022MTP}])\\
   \cline{1-3}
   \hhline{|===|}
   Material &Common materials, e.g., SiC.  & Special materials,   \\
    & &   e.g., Bi$_2$Se$_3$  \\
   \cline{1-3}
    Surface state  &  \multirow{2}*{Surface-charge-induced mode}   &  \multirow{2}*{Dirac plasmons}\\
  supported mode &  & \\
   \cline{1-3}
    Bulk state  &  \multirow{2}*{Surface phonon polariton}   &  Multiple phonon \\
      supported mode &  & polariton modes\\
   \cline{1-3}
    Origin of the   &  $\cdot$ Residual surface-localized    & $\cdot$ Surface states ﻿﻿  \\
    above modes & $~~$charges  & $~~$topologically protected\\   
       & $\cdot$ Bulk state & $\cdot$ Bulk states \\     
     \cline{1-3}
      Coupling inside   &  \multirow{2}*{weak}   & \multirow{2}*{significant}  \\
      the two states   &     &  \\
      \cline{1-3}
  Modulation    &  Match with Planck window   & Surface states couple\\
  mechanism    &   or not  & and compete with\\
         &      & the bulk states \\
      \cline{1-3}
   
  \end{tabular}
  \label{summary}
 \end{table}

\section{Conclusion}
In summary, we have studied the effect of residual charges on the NFRET between two nanoparticles and two planar slabs made of SiC. The residual charges on the nanoparticles and slabs result the common host object into an analog of the topological insulator and thus give rise to an additional surface-charge-induced resonant mode in the low-frequency infrared range. This mode provides a new heat flux channel enhancing the NFRET. For temperatures as low as 10 K as considered for the two slabs the Planck window matches well with the surface-charge-induced mode resonance resulting in a giant surface-charge-induced enhancement of three orders of magnitude at $d = 100\,{\rm nm}$. Consequently, for temperatures around room temperature as considered for the nanoparticles, the Planck window is far away from the low-frequency surface-charge-induced mode, which accounts for a much weaker residual charge effect on NFRET with an enhancement factor of 2.5. Furthermore, for high temperatures where the corresponding thermal Planck window is far away from the surface-charge-induced modes and the Fr\"{o}hlich resonance dominates the heat transfer rather the surface-charge-induced resonance, we find that the NFRET is rather inhibited than enhanced due to the surface-charge effect. These observations provide an alternative way to modulate NFRET between nanostructures composed of only common trivial materials by constructing the topological insulator analog through applying the residual charges. Since residual surface charges can exist in micro- and nano-systems like NEMS/MEMS, our results offer new insight on understanding and modulating heat transfer in nano-systems.

\section*{Acknowledgements}
The authors thank Prof. Achim Kittel and Dr. Shangyu Zhang for the fruitful discussions. The support of this work by the National Natural Science Foundation of China (No. 51976045; No. 52206081) is gratefully acknowledged. M.G.L. also thanks for support from the China Postdoctoral Science Foundation (2021M700991). S.-A.\ B.\ acknowledges support from Heisenberg Programme of the Deutsche Forschungsgemeinschaft (DFG, German Research Foundation) under the project No.\ 461632548.
\addcontentsline{toc}{section}{Acknowledgements}

The data that support the findings of this study are available from the corresponding author upon reasonable request.

\bibliography{RCE_clean}

\end{document}